# POWER-LAW DISTRIBUTIONS, THE H-INDEX, AND GOOGLE SCHOLAR (GS) CITATIONS: A TEST OF THEIR RELATIONSHIP WITH ECONOMICS NOBELISTS

By

Stephen J. Bensman
LSU Libraries
Louisiana State University
Baton Rouge, LA 70803 USA
E-mail: notsjb@lsu.edu

Alice Daugherty
LSU Libraries
Louisiana State University
Baton Rouge, LA 70803 USA
E-mail: adaugher@lsu.edu

Lawrence J. Smolinsky
Department of Mathematics
Louisiana State University
Baton Rouge, LA 70803 USA
E-mail: smolinsk@math.lsu.edu

Daniel S. Sage
Department of Mathematics
Louisiana State University
Baton Rouge, LA 70803 USA
E-mail: sage@math.lsu.edu

And

J. Sylvan Katz
SPRU, Jubilee Building
University of Sussex
Falmer, Brighton, BN1 9SL, UK
E-mail: j.s.katz@sussex.ac.uk

**Abstract**

This paper comprises an analysis of whether Google Scholar (GS) can construct documentary sets relevant for the evaluation of the works of researchers. The researchers analyzed were two samples of Nobelists in economics: an original sample of five laureates downloaded in September, 2011; and a validating sample of laureates downloaded in October, 2013. Two methods were utilized to conduct this analysis. The first is distributional. Here it is shown that the distributions of the laureates' works by total GS citations belong within the Lotkaian or power-law domain, whose major characteristic is asymptote or "tail" to the right. It also proves that this asymptote is conterminous with the laureates' h-indexes, which demarcate their core œuvre. This overlap is proof of both the ability of GS to form relevant documentary sets and the validity of the h-index. The second method is semantic. This method shows that the extreme outliers at the right tip of the tail—a signature feature of the economists' distributions—are not random events but related by subject to contributions to the discipline for which the laureates were awarded this prize. Another interesting finding is the important role played by working papers in the dissemination of new economic knowledge.

**Introduction**

This paper presents a test of the validity of the h-index and Google Scholar (GS) by analyzing the distribution of GS citations to the works of Nobelists in economics. In her authoritative review of informetrics Wilson (1999) noted that this discipline is governed by a series of mathematical laws whose commonality is that "the form of the distribution of numbers of items over the number of sources, when appropriately displayed, is variously described as extremely right-skewed, reversed-J-shaped, or quasi-hyperbolic" (p. 165). In informetric terms, a "source" is something like an author that produces an "item" like an article, and, when plotted on Cartesian coordinates with sources on the Y-axis and items on the X-axis, the distribution is characterized a long, horizontal asymptote or "tail" extending to the right. Distributions such as these have been hypothesized to have resulted from compound, contagious Poisson models that can be divided into two basic types: 1) those that can be logarithmically transformed to the normal distribution and belong to the "Gaussian domain"; and 2) those that cannot be so transformed and approximate inverse power laws belonging to the "Lotkaian domain". If the distribution of the GS citations to the works of the Nobelists form such a reversed J-curve and the h-index approximates the start of the asymptote or "tail," this will be considered a validation of not only of the h-index as demarcating the important œuvre of a given researcher but also of GS's ability to form relevant sets. The composition and authorship of the œuvres will be analyzed, and it will be determined whether the distributions belong to the Gaussian domain or the Lotkaian domain of the inverse power law. The distributions will also be semantically analyzed to test whether the documentary set is related by subject to the disciplinary contributions for which the laureates were awarded their prize.

**The Stochastic Models**

The 19$^{th}$ century was governed by what may be termed the normal paradigm. According to this paradigm, all distributions in nature and society conform to what came to known as the normal distribution. The normal distribution was originally developed contemporaneously by Pierre Simon Laplace and Carl Friedrich Gauss as a law of error for observations in astronomy and geodesy. It is also commonly called the "Gaussian distribution." Its major premise is that the arithmetic mean of a number of measurements or observed values of the same quantity is a good choice for the value of the magnitude of this quantity on the basis of the measurements or observations in hand, and it takes the form of the famous bell-shaped curve, where the three main measures of central tendency—mean, median, and mode—equal each other. Thus, by the normal law of error the arithmetic mean is by definition representative of the population.

The normal paradigm was destroyed by the British biometric school, which sought to place Darwin's theory of evolution on a firm mathematical basis. The first major blow against this paradigm was struck by the founder of this school, Francis Galton (1879), in a landmark paper entitled the "The Geometric Mean, in Vital and Social Statistics." In this paper Galton questioned whether the main assumptions of the normal law of error—that the arithmetic mean is the correct measure of central tendency and errors in excess and deficiency of this truth are equally probable—is applicable to vital and social statistics and instead proposed the geometric mean, which is calculated by first transforming the measures to the logarithmic scale, calculating the arithmetic mean of these transformations, and then taking the antilog of this arithmetic mean. In an accompanying paper entitled "The Law of the Geometric Mean," McAlister (1879) developed mathematically for the first time the lognormal distribution. In his treatise on probability Keynes (1921, pp. 198-199) paid special attention to the lognormal distribution, stating that it depends upon the assumption that the logarithms of the observations obey the normal law of error, making their arithmetic mean the most probable value of the logarithms of the quantity and the geometric mean of them as the most probable value of the quantity. In their monumental work on the lognormal distribution Aitchison and Brown (1957, pp. 1-2 and 100-106) describe examples of the lognormal distribution appearing in a broad array of fields and phenomena, stating that "the lognormal is as fundamental a distribution in statistics as is the normal, despite the stigma of the derivative nature of its name." According to them, it arises from a theory of elementary errors combined by a multiplicative process, just as the normal distribution arises from a theory of elementary errors combined by addition. From this it can be seen that the lognormal distribution falls within the Gaussian domain.

Galton did not understand the full ramifications of his insight, and the task of destroying the normal paradigm was accomplished by his protégé, Karl Pearson. In a series of papers on the mathematical theory of evolution, Pearson dealt with the problem of skew variation in homogeneous material, proving that the normal distribution was nonexistent in reality. In the place of an ubiquitous normal distribution, Pearson posited a system of skew frequency curves, of which the most important was Type III—chi-square or gamma—which could be unimodal with restricted range on the left and a long tail to the right (Bensman, 2000).

The British biometrician, who did the most to develop the stochastic models governing informetrics, was Pearson's student, George Udny Yule, who incorporated Pearson's findings into two models based upon the Poisson process, i.e., the random occurrence of events over time and space. The first was the negative binomial distribution, which was advanced by Greenwood and Yule (1920) in a study of accident rates of British female munitions workers during World War I. The Greenwood and Yule model can be explained simply in the following manner. Each female worker was considered as having a mean accident rate over a given period of time or her own lambda. Thus, the accident rate of each female worker was represented by a simple Poisson distribution. However, the various female workers had different underlying probabilities of having an accident and therefore different lambdas. Greenwood and Yule posited that these different lambdas were distributed in a skewed fashion described by Pearson's Type III or gamma distribution, and therefore certain workers had a much higher accident rate than the others and accounted for the bulk of the accidents. Greenwood and Yule attempted but failed to develop a usable contagious model, but this was done by Eggenberger and Pólya (1923) in a paper that analyzed the number of deaths from smallpox in Switzerland in the period 1877-1900. They derived their model off an urn scheme that involved drawing balls of two different colors from an urn and not only replacing a ball that was drawn but also adding to the urn a new ball of the same color. In this way numerous drawings of a given color increased the probability of that color being drawn and decreased the chance of the other color being drawn. According to Feller (1943), the Pólya-Eggenberger form was a product of "*true contagion*," because each favorable event increases (or decreases) the probability of future favorable events, while the Greenwood-Yule model represented "*apparent contagion*," since the events are strictly independent and the distribution is due to the probabilistic heterogeneity of the population. Given that Greenwood-Yule and Pólya-Eggenberger reached the negative binomial on different stochastic premises—the first on heterogeneity, the second on contagion—Feller posed the conundrum that one therefore does not know which process is operative when one finds the negative binomial. Here it should be emphasized that Feller considered that this conundrum was applicable not just to the negative binomial but generally to compound Poisson distributions that will always appear "contagious" (p. 398).

Yule (1925) laid the bases of the stochastic model underlying power-law distributions with another Poisson model that he developed to explain the evolutionary theory of John Christopher Willis on the distribution of species. To explain Willis' evolutionary theory in terms of mathematical statistics, he developed a Poisson model in which evolutionary saltations or mutations were treated as random events occurring over time. An important aspect of this model—perhaps the most important—is its ability to show cumulative effects over time. According to Yule's model, the number of species (i.e., size of a genus) increases in a geometric progression: if there is 1 species at time zero, there will be 2 species at time one, 4 species at time two, 8 species at time three, 16 species at time four, etc. This made "the doubling-period for species within the genus" (p. 25) the natural unit of time. Yule (1925) provided a generic demonstration of his model functioning over time with a hypothetical set of 1,000 genera operating over 6.28 doubling-periods (pp. 43-50). During this period, the range of the genera by size increased from a maximum of two genera with more than nine species to a maximum of 116 genera with more than 50 species, whereas the number of "monospecific" genera decreased from 571 to 343. As the process continued, the distribution increasingly approximated the negative exponential J-curve so common in informetrics.

Yule's model played no role in evolutionary theory, and its importance was not realized until it came to the attention of Herbert Simon (1955), winner of the 1978 Nobel Prize in economics, who gave it the name "the Yule distribution" in a seminal paper on skew distribution functions. In this paper, Simon (1955) states that his purpose is to analyze a class of distributions that appears in a wide range of empirical data, particularly data describing sociological, biological, and economic data. According to him, these distributions appear so frequently in phenomena so diverse that one is led to the conjecture that the only property they can have in common is a similarity in the structure of the underlying probability mechanisms. He describes these curves as "J-shaped, or at least highly skewed, with very long upper tails" (p. 425), i.e., negative exponential curves. It must be noted that Simon states that the stochastic processes are similar to those of the negative binomial but specifically rejects this as a suitable model (p. 426). He then introduces what he terms "the Yule distribution" (p. 426), based upon the incomplete beta function. In a book on skew distributions and the size of business firms, Ijiri and Simon (1977) give the following succinct, nonmathematical description of the stochastic process underlying the Yule distribution:

> This process starts with a few elements all of unit size in order to initialize the population. At each epoch, $\tau$, the aggregate size of the population (the sum of the sizes of the elements of the population) is increased by one unit. Two assumptions govern the rule on which element is to receive the unit increment. The first assumption states that there is a constant probability that the unit goes to a new element not previously in the population. If this happens, a new element of unit size is created in the population. If the unit does not go to a new element, it goes to an old element. In this case, the selection of the old element to which the unit is given is governed by the second assumption: the probability that the selected element is of size *i* (before the unit is added) is proportional to the aggregate size of all elements with size *i* (*i* times the number of elements with size *i*). The second assumption incorporates Gibrat's law of proportionality. Hence, big or small, each element has the same chance of growing at any given percentage in a given period. (p. 66)

Ijiri and Simon (1977) define the Gibrat law of proportionality as "the assumption that the expected percentage rate at which something will grow is independent of the size it has already attained," stating that "the Gibrat assumption virtually guarantees that the distribution will be highly skewed, with a long upper tail" (p. 24). Here it should be stated that the probability of a new unit going to an element of a given size *i* is proportional to the aggregate size of the *i* which stipulates that the Yule distribution is based upon a success-breeds-success mechanism, because the bigger and more numerous the elements of a given size *i,* the higher the probability of a new unit going to one of them. Simon (1955) showed that the Yule model provided good fits to empirical data sets not only of biological species but also city sizes, income distribution, word frequencies, and—most interestingly—papers authored by chemists, economists, mathematicians, and physicists. This paper is of historical importance for fitting the Yule-Simon model to empirical distributions of words by frequency and publications by scientists. It marked the first direct connection of a stochastic model to informetrics and its laws. From the perspective of this paper, it is important to note that Newman (2005, pp. 340-343 and 448) identified as one of the two main mechanisms of power laws "the Yule process"—"a rich-get-richer mechanism in which the most populous cities or best-selling books get more inhabitants or sales in proportion to the number they already have"—stating, "Yule and later Simon showed mathematically that this mechanism produces what is now called the Yule distribution, which follows a power law in its tail" (p. 348).

**Power Laws: Their Nature and Importance**

In his seminal paper on power laws Newman (2005) gives the following succinct definition and statement of importance of these distributions:

> When the probability of measuring a particular value of some quantity varies inversely as a power of that value, the quantity is said to follow a power law, also known variously as Zipf's law or the Pareto distribution. Power laws appear widely in physics, biology, earth and planetary sciences, economics and finance, computer science, demography and the social sciences. … (p. 324)

Power laws are of interest to this paper, because they have become accepted models for informetric laws and the distributional structure of the World Wide Web (WWW).

Informetric laws have long been hypothesized to be functions of compound Poisson or "contagious" distributions. One of the best examples of this approach was that of Bookstein (1990; 1997), who in a three-part article on informetric distributions rigorously analyzed three mathematically defined informetric laws—Lotka's Law, Bradford's Law, and Zipf's Law—together with Pareto's Law on Income, concluding that these laws are variants of a single distribution despite marked differences in their appearance and locating this distribution within the family of compound Poisson distributions. The negative binomial is one example of such a distribution, and Simon posited that what he termed "the Yule distribution" resulted from the same stochastic processes as the negative binomial with the concurrence of Newman.

The power-law approach to informetric laws is cogently presented by Egghe (2005) in his groundbreaking book on "Lotkaian informetrics." In this book Egghe focuses on "two-dimensional informetrics," where "one considers sources (e.g. journals, authors, …) and items (being produced by a source - e.g. articles) and their interrelations." He then states:

> … Essentially in two-dimensional informetrics the link between sources and items is described by two possible functions: a size-frequency function f and a rank-frequency function g. Although one function can be derived from the other, they are different descriptions of two-dimensionality. A size-frequency function f describes the number f(n) of sources with n = 1, 2, 3, … items while a rank-frequency function g describes the number g(r) of items in the source on rank r = 1, 2, 3, … (where the sources are ranked in decreasing order of the number of items they contain (or produce)). So, in essence, in f and g, the role of sources and items are interchanged. This is called duality, hence f and g can be considered as dual functions. (p. 1)

Egghe classifies Pareto's and Zipf's laws as based on the rank-frequency function, whereas he states that Lotka's law is based on the size-frequency function, where $f(n) = C / n^a$, where $C > 0$ and $a \geq 0$, implying that the function f decreases. He declares that the only axiom used in his book is that the size-frequency function is Lotkaian, hence his term "Lotkaian informetrics." In this paper we will follow Egghe and base our power-law analysis on Lotka's law and the size-frequency function. However, as an interesting aside, Egghe (2005, pp. 22-24) includes Bradford's Law of Scattering among the distributions compatible with Lotkaian informetrics, confirming Bookstein's view of the stochastic similarity of this law to Lotka's Law, Zipf's Law, and Pareto's Law. Bradford's Law is historically the most important of the informetric laws and is based upon the rank-frequency function.

The power-law model of the World Wide Web (WWW) is most associated with the name of Albert-László Barabási. In developing his model Barabási highlighted key characteristics of power-law distributions. The essence of this model is the concept of "scale-free network" or a network whose degree distribution follows a power law at least asymptotically. In a seminal paper Barabási and Albert (1999) utilized graph theory to analyze the structure of the Web. According to this theory, a network is comprised of vertices or nodes or elements of the system and edges representing interactions between them. In Web terminology, vertices are virtual documents, and edges are the hyperlinks connecting them.  Barabási and Albert (1999) started with the hypothesis that the Web followed the Erdős-Rényi random model.  Barabási and Bonabeau (2003) summed up what the results should have been with this model thus:

> An important prediction of random-network theory is that, despite the random placement of links, the resulting system will be deeply democratic: most nodes will have approximately the same number of links. Indeed, in a random network the nodes follow a Poisson distribution with a bell shape, and it is extremely rare to find nodes that have significantly more or fewer links than the average.  Random networks are also called exponential, because the probability that a node is connected to k other sites decreases exponentially for large k.  (p. 52)

Instead Barabási and Albert (1999) found that the probability $P(k)$ that a vertex in the Web interacts with k other vertices decays according to the following power law: $P(k) \sim k^{-e}$, where e is a constant parameter of the distribution known as the exponent or scaling parameter.  Barabási and Albert (1999) located the inability of the random model to predict Web reality in its failure to incorporate two generic mechanisms of real systems: (i) networks expand continuously by the addition of new vertices, and (ii) new vertices attach preferentially to sites that are already well connected, resulting in a "rich-get-richer" phenomenon (p. 37).  Albert, Jeong, and Barabási (1999) in a paper published almost simultaneously with the original presentation of the model pointed out two important implications of the asymptote or power-law tail. First, this tail indicates that the probability of finding documents with a large number of links is significant as the network connectivity is dominated by highly connected Web pages, and this particularly applies to inlinks, making the probability of finding very popular addresses, to which a large number of other documents point, non-negligible.  Second, the dominance of connectivity by a relatively small number of nodes made the Web a small-world network despite its enormous and exponentially growing size.  Defining the "diameter" of the Web as the number of clicks it would take to move from one randomly selected document to any other, the authors calculated that this diameter was only 19 that would increase to merely 21 with a 1000% increase in Web size.  As will be seen, these findings have important implications for the successful operation of the Google search engine.

In a popularized presentation of the model Barabási and Bonabeau (2003) utilized the term "scale-free" to explain it.  The term "scale-free" is a synonym for "scale-invariant" and "self-similar"—concepts understood in mathematics and physics for a long time.  Barabási and Bonabeau (2003) stated that, when they began to map the Web, they expected that the nodes would follow bell-shaped distributions like the normal law of error as do people's heights but instead found the Web dominated by a relatively small number of nodes that are connected to many other sites.  Such nodes were impossible under the random model, and it was as if they had stumbled upon a significant number of people who were 100 feet tall.  This prompted them to call the Web distribution "scale-free" in the sense that some nodes have a seemingly unlimited number of links and no node is typical of the others.  In other words, any measure of central tendency such as the arithmetic mean is meaningless.

Barabási's findings were corroborated by Huberman (2001), who codified his research in an important book entitled *The Laws of the Web*.  Huberman found the same skewed distributions as did Barabási, hypothesizing also that these resulted from the Web being governed by a power law, stating, for example, that the probability of finding a Web site with a given number of pages, n, is proportional to $1/n^\beta$, where $\beta$ is a number greater than or equal to 1.   Seeking an analogy for the power-law structure of the Web, Huberman (2001, pp. 30-31) referred to Zipf's Law.  Of great interest in this book is Huberman's further explanation of what is meant by a distribution resulting from a power law being "scale-free." Huberman presented this in the following layperson terms:

> The interesting thing about a distribution with a power law form is that if a system obeys it, then it looks the same at all length scales. What this this means is that if one were to look at the distribution of site sizes for one arbitrary range, say just sites that have between 10,000 and 20,000 pages, it would look the same as that for a different range, say from10 to 100 pages. In other words, zooming in or out in the scale at which one studies the Web, one keeps obtaining the same result…. (p.25).

Newman (2005) concurred with this definition of "scale-free," stating that "a power law is the only distribution that is the same whatever scale we look at it on" (p. 334).

**Power Law Distributions: Their Mathematics and Fitting**

As stated above, in this paper we will follow Egghe (2005) in basing our power-law analysis on Lotka's Law in an implementation of his "Lotkaian informetrics" and adopting his terminology. Lotka's Law is the eponymic name of a law of scientific productivity derived by Alfred J. Lotka. This was the first informetric or scientometric power law, and this law is often termed "the inverse square law of scientific productivity." To obtain the data for deriving his law, Lotka (1926) made a count of the number of personal names in the 1907-1916 decennial index of *Chemical Abstracts* against which there appeared 1, 2, 3, etc. entries, covering only the letters A and B of the alphabet. He also applied a similar process to the name index in Felix Auerbach's *Geschichtstafeln der Physik* (Leipzig: J. A. Barth, 1910), which dealt with the entire range of the history of physics through 1900. By using the latter source, Lotka hoped to take into account not only the volume of production but also quality, since it listed only the outstanding contributions in physics. In informetric terms the scientists are the "sources" demarcated on the Y-axis, and their contributions or mentions are the "items" demarcated on the X-axis. On the basis of this data, Lotka derived what he termed an "inverse square law", according to which of any set of authors, ca. 60% produce one paper, whereas the percent producing 2 is $1/2^2$ or ca. 25%, the percent producing 3 equals $1/3^2$ or ca. 11.1%, the percent producing 4 is $1/4^2$ or ca. 6.3%, etc. Thus, of 1000 authors, 600 produce 1 paper, 250 produce 2 papers, 111 produce 3 papers, and 63 produce 4 papers.

To explain the mathematics of Lotka's Law, Egghe (2005, pp. 14-16) provides the following equation for this law:

$$F(n) = C/n^a$$

where $C > 0$ and $a \geq 0$ are constants. Of these two constants the exponent a—named "the Lotka exponent" by Egghe—is the most important one as it indicates the degree of concentration and complexity. C is only used to make sure that $\Sigma f(n)$ gives the total number of sources. The size of the Lotka exponent is one way to judge whether a power-law is operational, and for this one needs to know various demarcation points. As a starting point for this, one can use the fact that Lotka (1926, p. 320) calculated an exponent of 1.888 for his *Chemical Abstracts* data and 2.021 for his Auerbach data—both close to 2 and hence the term "inverse square law." Egghe collaborated with Ronald Rousseau in writing his book, and Rousseau and his son (Rousseau and Rousseau, 2000) constructed a computer program named LOTKA, which they designed to fit power-law distributions to observed data. To do this, they had to determine limits for the Lotka exponent, which we will use as demarcation points. Surveying the previous literature, they noted that Egghe (1990) had mathematically proven that a Lotka exponent above 3.0 cannot occur. Rousseau and Rousseau (2000) also noted that another method usually yielded exponents above 1.7. As a result they set of the limits of this exponent in their LOTKA program at 1.26 and 3.49, figuring these limits would enough to take care of most eventualities. In his book Egghe (2005, p. 15) points out that it has been emphasized in the literature that the statistical fitting of a power law often is impossible, certainly for low Lotka exponents. In a discussion on SIGMETRICS of the impact factor, which is based upon the arithmetic mean, Newman (2011, Sept. 1) posted the following commentary that is of great importance for this matter:

> The crucial issue here is the point…that for the Central Limit Theorem to be applicable, and hence for the mean to be valid, the distribution has to fall in the "domain of attraction of the Gaussian distribution". As others have pointed out, the Pareto or power-law distribution to which the citation distribution is believed to approximate, does not fall in this domain of attraction if its exponent is less than 3.

As a result of the above analysis, three statements can be made about the demarcation points for the Lotka exponent: 1) exponents close to 2 are most indicative of a power law; 2) exponents much below 2 are probably too erratic to be indicative; and 3) with exponents $\geq 3$ one is transcending from the Lotkaian to the Gaussian domain. Here an important fact must be noted. Whereas power laws are not within the Gaussian domain of attraction, the negative binomial is. Thus, in a statistical manual widely used by ecologists, Elliot (1977, pp. 30-36) stipulates that the negative binomial be logarithmically transformed—i.e., converted to the lognormal—to satisfy the assumption of the normal distribution associated with many parametric statistical techniques such as the Pearson correlation.

An extremely important and influential paper on the statistical fitting of power-law distributions to empirical data has recently been published by Clauset, Shalizi, and Newman (2009). In this paper they note that the Lotka exponent typically lies in the range $2 < a < 3$ (p. 662), but they state that the detection and characterization of power laws is complicated by two basic problems: 1) the difficulty of identifying the range over which power-law behavior holds; and 2) complicated by the large fluctuations that occur in the tail of the distribution—the part of the distribution representing large but rare events (p. 661). We will discuss their methods and findings, but first it is necessary clarify one thing. Clauset, Shalizi, and Newman (2009) base their analysis on MatLab algorithms, and MatLab has a definition of the "lognormal distribution" that differs subtly from the traditional one. Thus, the MatLab documentation (MathWorks Documentation Center, 2014) defines the log-normal distribution thus:

> The lognormal distribution is a probability distribution whose logarithm has a normal distribution…. [It] is closely related to the normal distribution. If x is distributed lognormally with parameters μ and σ, then log(x) is distributed normally with mean μ and standard deviation σ…..

Historically the lognormal has been defined as a distribution which has already been logarithmically transformed and whose logarithms fit the normal distribution; whereas MatLab defines a lognormal distribution as a distribution that has not been logarithmically transformed but whose logarithms would fit the normal distribution once the transformation had been implemented.

In respect to the first problem Clauset, Shalizi, and Newman (2009) summarize the difficulty thus:

> …it is normally the case that empirical data, if they follow a power-law distribution at all, do so only for values of x above some lower bound $x_{min}$. Before calculating our estimate of the [Lotka exponent] scaling parameter a, therefore, we need to first discard all samples below this point so that we are left with only those for which the power-law model is valid. Thus, if we wish our estimate of α to be accurate, we will also need an accurate method for estimating $x_{min}$. If we choose too low a value for $x_{min}$, we will get a biased estimate of the [Lotka exponent] since we will be attempting to fit a power-law model to non-power-law data. On the other hand, if we choose too high a value for $x_{min}$, we are effectively throwing away legitimate data points $x_i < \dot{x}_{min}$, which increases both the statistical error on the [Lotka exponent] and the bias from finite size effects. (p. 669)

Thus, power-laws are only operative on the asymptote above some $x_{min}$. The difficulty is compounded by the fact that the distributional structure of the asymptote can be enormously complex. In an extremely interesting study of collaborative patterns in nanoscience Milojević (2010a) exposed the difficulties of fitting empirical data to a supposed power-law distribution over its full range. She found three collaboration modes that corresponded to three distinct ranges in the distribution of collaborators: (1) for authors with fewer than 20 collaborators (the majority) preferential attachment did not hold and they form a lognormal "hook" instead of a power law; (2) authors with more than 20 collaborators benefit from preferential attachment and form a power law tail; and (3) authors with between 250 and 800 collaborators are more frequent than expected because of the hyperauthorship practices in certain subfields.

The third part of the finding by Milojević is an example of the second major difficulty pointed out by Clauset, Shalizi, and Newman for fitting empirical data to power-law distributions--large fluctuations on the extreme right of the tail of the distribution representing large but rare events.

Clauset, Shalizi, and Newman (2009, p. 663) note that power-law distributions are either continuous, governing continuous real numbers, or discrete, where the quantity of interest is only a discrete set of values, typically positive integers, and they conduct their analysis accordingly. Since we are dealing with citation counts, we will restrict ourselves to their findings in respect to discrete data. Clauset, Shalizi, and Newman (2009) tested 10 sets of discrete data for power-law behavior against the five other following distributions: Poisson; log-normal; exponential, stretched exponential or Weibull;

and power-law + cut-off. They found only one dataset—the frequency of occurrence of unique words in the novel *Moby Dick* by Herman Melville—that provided good support for the power-law distribution and one dataset that provided no support. Two other datasets provided moderate support, but—significantly—four other datasets, including one concerning the Internet, provided support for the power-law distribution on condition of the implementation of the cut-off. Clauset, Shalizi, and Newman (2009) summarized their findings thus:

> …For many [datasets] the power-law hypothesis turns out to be, statistically speaking, a reasonable description of the data. That is, the data are compatible with the hypothesis that they are drawn from a power-law distribution, although they are often compatible with other distributions as well, such as log-normal or stretched exponential distributions. In the remaining cases the power-law hypothesis is found to be incompatible with the observed data. In some instances…the power law is plausible only if one assumes an exponential cut-off that modifies the extreme tail of the distribution. (p. 690)

It will be seen that the problem with the extreme tail of the distribution is a particular problem with GS citations to the works of Nobelists in economics as these tend to concentrate heavily on a few key works of the prize winners, making the right tip of the tail very messy. The possible compatibility with the lognormal distribution is also of interest. Thus, Mitzenmacher (2004) noted that power-law distributions were pervasive in computer science and that the conventional wisdom in this discipline was that file sizes are governed by power-law distributions. He found that power-law distributions and lognormal distributions connect quite naturally and that lognormal distributions have arisen as a possible alternative to power law distributions across many fields.

**Google Scholar and the H-Index: The Need for Left Truncation**

  Bensman (2013) analyzed the theoretical bases of the Google search engine, which operates on an algorithm named PageRank after Larry Page, one of the founders of Google. He found that the basic premise, on which PageRank operates, is the same as the one on which Eugene Garfield based his theory of citation indexing that underlies the *Science Citation Index*, which he created, i.e., that subject sets of relevant documents are defined semantically better by linkages than by words. Google incorporated this premise into PageRank, amending it with the addition of the citation influence method developed by Francis Narin and the staff of Computer Horizons, Inc. (CHI). This method weighted more heavily citations from documents which themselves were more heavily cited. Bensman noted that PageRank can be considered a further implementation of Garfield's theory of citation indexing at a higher technical level, and, analyzing GS citations to the works of Nobelists in chemistry, he obtained results compatible with Garfield's Law of Concentration—i.e., that the scientific journal system is dominated by a small, interdisciplinary core of highly cited titles—and his view on the importance of review articles. It must be emphasized that Bensman (2013) did not conduct his analysis by PageRank but in terms of total Google Scholar (GS) citations—the equivalent of how Garfield did his analysis except the citations were not references from journal articles but inlinks from Web sites. The same will be done in this paper. However, that which follows pertains to the probabilistic milieu in which the Google search engine operates. First, the importance of a Web page is directly related to the steady-state probability that a random Web surfer ends at the page after following a large number of links, and there is a larger probability that the surfer will wind up at an important page (i.e., one with many cites to it) than at an unimportant page. It is easily seen how this process can be expedited by Barabási's theory of the power-law structure of the WWW. Second, web search engines operate on what is known as the "Probability Ranking Principle," whereby they respond to queries by ranking retrieved documents in order of decreasing probability of relevance.  Given the structure of the WWW, the drop in relevance and the concomitant increase in irrelevance or error is exponential.

As a result of the probabilistic milieu in which the Google search engine operates, one cannot obtain full distributions of GS citations, and there is a need for a logical point of left truncation.  In evaluating publications of researchers, a logical point of such truncation—or $x_{min}$ in the terminology of Clauset, Shalizi, and Newman—is the h-index, which has become one of the leading measures for evaluating scientists, scientific institutions, and scientific publications.  It was created by J. E. Hirsch, a physicist at the University of California at San Diego.  In his initial formulation Hirsch (2005) defined his h-index thus: "A scientist has index h if h of his or her Np papers have at least h citations each and the other (Np - h) papers have ≤ h citations each" (p. 16569), where Np is the number of papers published over n years.   The h-index has certain advantages for analyzing power-law distributions derived from WWW data as it controls for both ends of the distribution.  It discounts for the extreme outliers on the right, and it determines a possible point of truncation on the left—i.e., a $x_{min}$—below which the data tends to become censored, irrelevant, and not worth analyzing.  In its structure the h-index resembles Lotka's law, and it suffers from the same major defect.  Egghe (2005) pinpointed this defect by stating that, of the many information production processes, it is the only one where items can have multiple sources.  In a follow-up paper to his original one, Hirsch (2007) began to confront this problem by modifying his initial formulation in the following important way:  "The h index of a researcher is the number of papers **coauthored** [emphasis added] by the researcher with at least h citations each" (p. 19193). Having made this important modification, Hirsch (2010) openly acknowledged in a following paper that "the most important shortcoming of the h-index is that it does not take into account in any way the number of coauthors of each paper," stating, "This can lead to serious distortions in comparing individuals with very different coauthorship patterns…." (p. 742).  This is a particularly important problem when utilizing Google Scholar (GS), which retrieves works to which researchers have contributed no matter what the position of the researcher in the authorship structure—primary author, secondary author, or even editor in the case of books.

**Data Analysis: Creation and Validation of the Documentary Sets**

To test the validity of Google Scholar (GS) for evaluating the contribution of economists to their field, GS cites to the works of two groups of Nobel laureates in economics were downloaded from Web with the aid of the Publish or Perish (PoP) computer program developed by Anne-Wil Harzing (2010) that is freely available at http://www.harzing.com. This program establishes both bibliographic and statistical control over GS, and Jacsó (2009)—perhaps the most persistent critic of GS—found this software to be "a swift and elegant tool to provide the essential output features that Google Scholar does not offer" (p. 1189). The first group and the year of their prize were the following: Harry Markowitz (1990), James Heckman (2000), Paul Krugman (2008), Elinor Ostrom (2009), and Peter Diamond (2010). Data for this group of economists were downloaded in September, 2011. The second group of laureates comprised the following: Robert William Fogel (1993), C. W. J. Granger (2003), Thomas J. Sargent (2011), Lloyd S. Shapley (2012), and Eugene F. Fama (2013). Data for this second group of laureates were downloaded October, 2013. The purpose of the second group was to validate—or invalidate—the findings made with the analysis of the first group. The temporal spacing of both groups was the same: the earliest, middle, and most recent prize years were 10 years apart, whereas the latest three prizes were one year apart. By this it was hoped to better understand the effect of time.

Harzing (2013; 2014) herself downloaded the data for the first group of laureates and graciously gave it to the primary author of this paper. She temporally spaced the prizes as above, and in the years with multiple winners, she selected the first Nobel Prize winner unless this Nobelist had a particularly common name that would cause a major homonym problem. The PoP program ranks the publications in descending order by GS cites and automatically calculates the h-index of an author. Harzing concentrated on validating only these h-index publications. These were verified individually to ensure they were published by the Nobelist in question. Any publications with substantial stray records were merged, especially if they were on the h-index threshold. The merging process did not substantially change the h-index, which in most cases remained the same, moving up or down by one in the others. Due to these procedures, her distributions were bifurcated: an upper, accurate range demarcated by the h-index; and a lower range increasingly full of error and irrelevancy as the number of cites per publication decreased.   The same procedures were followed in constructing the documentary sets for the second, validating sample of laureates with the same bifurcated distributions.

> Harzing (2008) justified truncation on the left at the h-index in the following terms:
> The advantage of the h-index is that it combines an assessment of both quantity (number of papers) and quality (impact, or citations to these papers)…. An academic cannot have a high h-index without publishing a substantial number of papers.  However, this is not enough. These papers need to be cited by other academics in order to count for the h-index.

According to her, the h-index is preferable over the total number of citations as it corrects for academics who might have authored a few highly-cited papers, but have not shown a sustained and durable academic performance, and it is also preferable over the number of papers as it corrects for uncited papers, thereby favoring academics who publish a continuous stream of papers with lasting and above-average impact.  Burrell (2007) concurred with Harzing's assessment of the h-index, defining its purpose as seeking to identify "the most productive core of an author's output in terms of most received citations," which he termed "the Hirsch core, or *h*-core" (p. 170).

Harzing (2013) subjected the GS cites to her sample of economists to intensive analysis, comparing them to GS cites to laureates in chemistry, medicine, and physics.  Her main findings can be summarized as follows.  First (p. 1070), the main difference in Web of Science and Google Scholar citations for Nobelists in economics from those in the other disciplines was largely due to their book publications.  And this was despite the recent introduction of the *Book Citation Index*. Whereas for the prize winners in the sciences and medicine, the top 10 publications were nearly always journal articles, often in prestigious journals such as *Nature*, *Science*, *Cell*, *The Lancet*, and *Physical Review.* Four of the five economics laureates had books amongst their top-10 publications, and all of these had a book in their top-3 most cited publications. For Ostrom all top-3 publications were books, whereas for Krugman and Markowitz two out of the top-3 publications were books. Second (p. 1071), the relationship between WoS and GS citations for the top-20 articles was extremely different for the economists than for the laureates in the other disciplines. The economists had between three and five times as many citations for their top-20 articles in GS than in WoS, but this difference is dwarfed by the difference for their entire citation record, which shows four to twelve times as many citations in GS, indicating economists benefit not only from the larger number of citing journal articles in GS but also from GS's extended coverage of books, book chapters, conference papers, and working papers. And, third, Harzing (2013, p. 1072) compared the GS and WoS rankings by both h-index and total cites of all the laureates within their disciplines and found these to be virtually identical. The ranking was completely identical for Medicine, whereas for the three other disciplines numbers 1 and 2 were swapped between the two databases with the result that the correlation between the two rankings was very high at 0.93. As noted above, Harzing's findings on this third point have been validated Bensman (2013), who utilized Harzing's GS data on the chemistry laureates to prove that the Google search engine is actually based upon a further implementation of Garfield's theory of citation indexing at a higher technical level, yielding results fully in conformance Garfield's theories on the structure of the scientific journal system.

**Data Analysis: The Statistical Tests**

The statistical tests will be based on the construction and analysis of histograms, whose vertical Y-axis will demarcate sources (works) and whose horizontal X-axis will demarcate items (GS cites). These citations will be grouped into 25 bins each covering 4% of the citation range. The 4%-bins will be assigned ordinal numbers from $1^{st}$ for the lowest to $25^{th}$ for the highest. There will be two such histograms for each laureate.  The first will be the full distribution or everything downloaded from GS by the PoP program.  As described above, this will divided into two parts: an upper, cleansed part above the h-index; and a lower part below the h-index full of irrelevance and error.  The full distribution will be inspected to determine if it has a horizontal asymptote—or "heavy tail"—and, if so, where approximately this asymptote begins.  If, for example, the h-index and the beginning of the horizontal asymptote both occur in the 1$^{st}$ 4%-bin, this can be considered an indication of three things: 1) the h-index and $x_{min}$ approximate each other; 2) the h-index is demarcating the core oeuvre of the researcher; and 3) GS has defined a relevant set.

Upon the completion of the above analysis, the distribution will be truncated on the left at the h-index, and a new histogram will be constructed on the same principles as above—with the citations along the entire range on the horizontal x-axis grouped 25 4%-bins numbered from 1$^{st}$ to 25$^{th}$.  This histogram will inspected to see if it has the tell-tale sign of a power-law distribution—a negative j-curve with a horizontal asymptote or "heavy tail" of the same general shape as the full distribution.  The truncated distribution will then be subjected to the following tests.  First, there will be performed the index of dispersion test to determine whether the distribution is random, regular, or contagious resulting from some cumulative advantage process as preferential attachment.  Second, the MatLab techniques developed by Clauset, on which Clauset, Shalizi, and Newman (2009) based their analyses, will be utilized to estimate the Lotkaian exponent and determine whether the distribution fits a power-law one.  In making these tests, the h-index will serve as the $x_{min}$  Here there must be emphasized an important limitation of employing the h-index as $x_{min}$ in power-law analyses.  Clauset, Shalizi, and Newman (2009) spell out this limitation thus:

> …Our experience suggests that n ≥ 50 is a reasonable rule of thumb for extracting
>
> reliable parameter estimates…. Data sets smaller than this should be treated with

> caution. Note, however, that there are more important reasons to treat small data sets with caution. Namely, it is difficult to rule out alternative fits to such data, even when they are truly power-law distributed, and conversely the power-law form may appear to be a good fit even when the data are drawn from a non-power-law distribution. (p. 669)

Thus, the analyses are only reliable with researchers with h-indexes of 50 or above, and this may not only affect individual researchers but rule out entire disciplines.

For the final two statistical tests, the h-index truncated range of the GS cites on the X-axis will be logarithmically transformed, and two histograms will be constructed by dividing this logarithmic range into 25 4%-bins. Although the logarithmic bins are equal, the effect of doing this is that in terms of the untransformed data the bins are exponentially increasing in width as one goes up the range, consolidating the "messy" observations at the extreme right of the asymptote. With the first histogram the X-axis will be on the logarithmic scale, but Y-axis will not. The purpose of this is so that histogram graphs the frequency counts of works in the logarithmic bins to determine the basic shape of the distribution. This histogram provides the basis of estimating whether the distribution falls within the Gaussian domain. If the curve approximates the signature bell-shape of the normal law of error, it is a sign that we are dealing with a lognormal distribution and are in the Gaussian domain. This assumption will be statistically verified by the Shapiro-Wilks test for normality. With the second histogram both the X-axis and the Y-axis will be on the logarithmic scale. The purpose of this histogram is to provide the basis for estimating whether the distribution is in the Lotkaian domain. According to Clauset, Shalizi, and Newman (2009), a standard way to test for a power-law distribution has been the following:

> …A common way to probe for power-law behavior, therefore, is to measure the quantity of interest x, construct a histogram representing its frequency distribution, and plot that histogram on doubly logarithmic axes. If in so doing one discovers a distribution that falls approximately on a straight line, then one can, if feeling particularly bold, assert that the distribution follows a power law, with a scaling parameter α [i.e., Lotka exponent ] given by the absolute slope of the straight line. Typically this slope is extracted by performing a least-squares linear regression on the logarithm of the histogram. This procedure dates back to Pareto's work on the distribution of wealth at the close of the 19$^{th}$ century. (p. 665)

Clauset, Shalizi, and Newman (2009, p. 689) warn that identifying power-law distributions by the approximately straight-line behavior of a histogram on a doubly logarithmic plot should not be trusted but do state that such straight-line behavior is a necessary condition for true power-law behavior. For this reason we will perform this test on the h-index truncated GS distributions of the laureates, and, if these distributions manifest such a behavior, we will consider it is a sign that we are dealing with the Lotkaian domain. In a subsequent paper we will show that restricting the use of this procedure to power-law identification is a misunderstanding of the purpose and utility of this test. This paper will also show that this test connects Lotka's law with the Yule distribution, providing the historical basis for Egghe's "Lotkaian informetrics."

**Data Analysis: The Paul Krugman Example**

The above statistical tests will be applied to both the original set of Nobelists—Harry Markowitz (1990), James Heckman (2000), Paul Krugman (2008), Elinor Ostrom (2009), Peter Diamond (2010)—downloaded in September, 2011, and the validating set of Nobelists--Robert William Fogel (1993), C. W. J. Granger (2003), Thomas J. Sargent (2011), Lloyd S. Shapley (2012), Eugene F. Fama (2013)--downloaded October, 2013. Each set of results will be combined into one table for comparative purposes. However, the tests will first be demonstrated and explained with the results for Paul Krugman. From the perspective of this paper, Krugman is the most interesting for the following three reasons. First, of all Nobel prizes the one in economics probably has the most societal impact, and Krugman's career reflects this as he has at least three jobs: professor of economics and international affairs at Princeton University; Centenary Professor at the London School of Economics; and his best-known job as an op-ed columnist for *The New York Times*, resulting in his being called "the most important political columnist in America" (Nobelprize.org, 2014). He is now reportedly being paid $225,000 by the City University of New York to lecture on "income inequality." Second, in an op-ed piece entitled "Open Science and the Econoblogosphere" Krugman (2012, Jan. 12) stated that the traditional method of submitting, being refereed, and then published in leading journals broke down in economics as far back in the 1980s as being too slow. Even in those days, according to him, nobody at a top school learned things by reading the journals; it was all done principally through National Bureau of Economic Research Working Papers (yellowjackets) being exchanged among members of informal working groups. Journals served as tombstones, good for only tenure committees, and the notion of journals as gatekeepers was largely fictional already 25 years ago. Krugman states that the system nowadays still works the same way except that the printed working papers have been replaced by rapid-fire exchange via blogs and online working papers, suggesting that GS cites may perhaps be a better measure of economists' importance than WoS ones. And, finally, Krugman was awarded the prize "for his analysis of trade patterns and location of economic activity," and his contribution to economics as identified as: "Integrated the previously disparate research fields into a new, international trade and economic geography" (Nobelprize.org, 2014). In this work Krugman (1996a; 1996b, pp. 44-46 and 92-97) confronted the problem of using power-law models to analyze the distribution of U.S. cities by size. He found that the size distribution of U.S. cities is startlingly well described by a simpler power law, i.e., the number of cities whose population exceeds S is proportional to 1/S. For him this simple regularity
27

was puzzling especially as it had apparently remained true for at least the past century. According to Krugman, standard models of urban systems offer no explanation of the power law, but the Yule-Simon model was the best try to date. However, he stated, while it can explain a power law, it cannot reproduce one with the right exponent, leaving us in the frustrating position of having a striking empirical regularity with no good mathematical theory to account for it. For these reasons Krugman (1996b) considered the Yule-Simon distribution not a model but a "story" (p. 44), and he stated that, while he had no resolution of the problem, he was convinced of three things: 1) the power-law model on city sizes is very real and tells us something very important about our economy; 2) some kind of random growth process Is the most likely explanation; and 3) the Yule-Simon model is so elegant an approach that it is the best game in town (p. 97). For the above reasons Krugman will be used as a model to demonstrate the statistical tests. His only drawback is that, unlike the other economists, Harzing did not achieve a full download on Krugman, and his full distribution is lacking the zero and one citation classes, but this has little effect on the analysis.

Figure 1 below is the histogram of Krugman's full distribution (lacking the zero and one citation classes), and it has the classic shape of a power-law distribution with a long asymptote—or tail--to the right. Two things are of great interest in this histogram. First, both the h-index and the start of the tail are located in the 1$^{st}$ or lowest 4% bin, indicating a close relationship of the h-index to $x_{min}$. Second, there is a scatter of four publications on the extreme right, making the tail extremely messy at its tip. Table 1 below sets forth statistics underlying this histogram. It states that Krugman had an h-index of 114—sufficiently above the required n of 50 for an adequate sample—and it summarizes the statistical consequences of this distribution. Although the h-index works account for only 12.0% of the 950 "works" in the full distribution, they account for 98.5% of the full GS citation range of 7,220 and 85.8% of the total of 93,827 GS citations. From the viewpoint of descriptive statistics, the h-index appears to have defined Krugman's core œuvre.

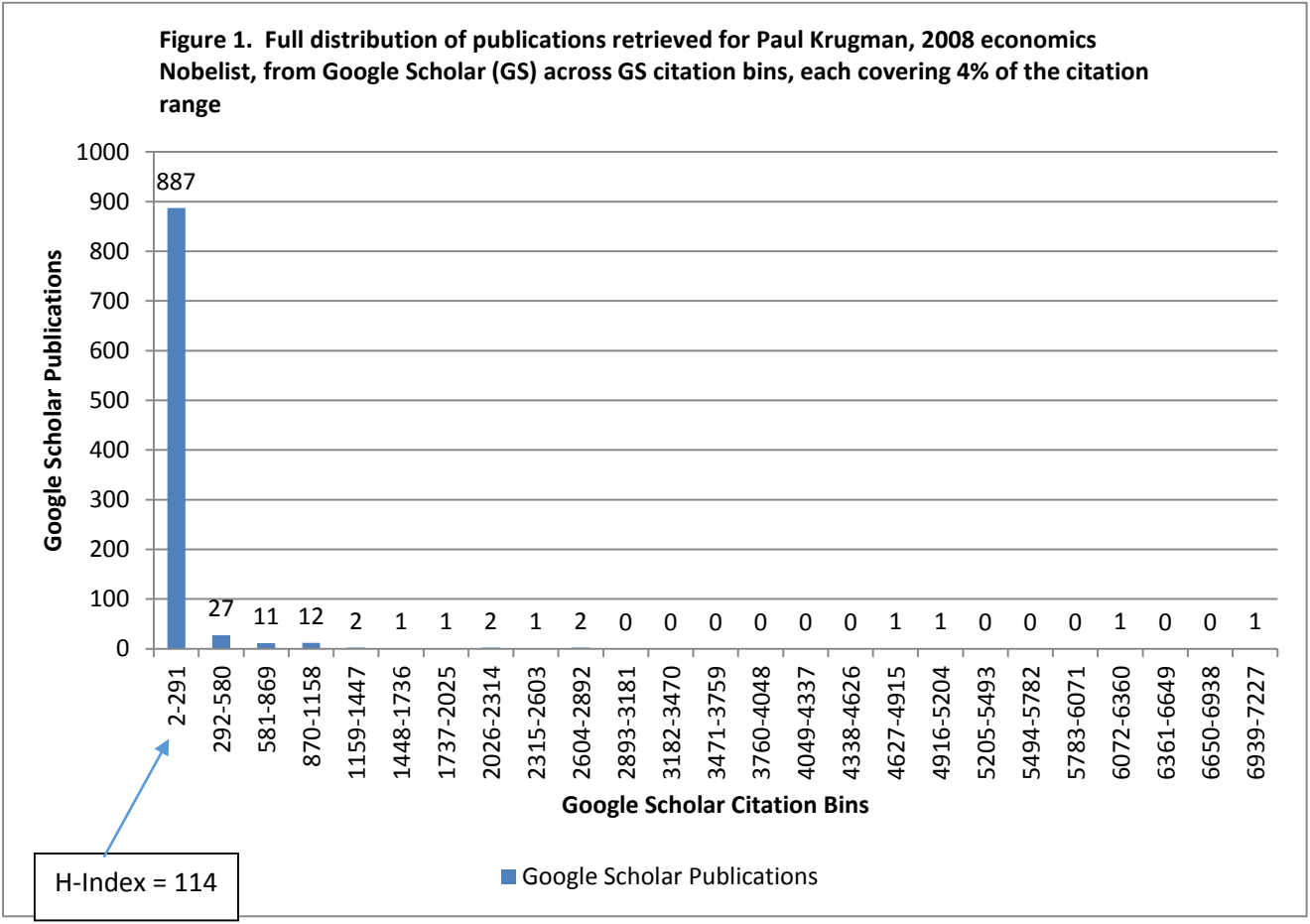

Figure 1. Full distribution of publications retrieved for Paul Krugman, 2008 economics Nobelist, from Google Scholar (GS) across GS citation bins, each covering 4% of the citation range

Given the operation of the Google search engine and Harzing's cleansing of the data, most of Krugman's supposed "works" below the h-index can be considered incoherent and irrelevant clutter. Figure 2 below is a histogram of the h-index truncated distribution of GS citations to Krugman's works constructed in the same way as the full distribution. It has the same shape as the full distribution with the same long and messy tail. Table 2 below summarizes the statistical characteristics of the h-index truncated distribution. Of greatest interest here are the results of the chi-squared index of dispersion tests for the Poisson distribution to determine whether these distributions were random or contagious. A full explication of this test is given in Elliott (1977, pp. 40-44). The basis of this test is that a key feature of the Poisson distribution is the following identity: lambda (Λ) = mean = variance.  Therefore, a comparison of the variance of a distribution to its mean is actually a comparison of its actual variance to its theoretical Poisson variance.  As a result, an index of dispersion equaling 1 is indicative of the Poisson.  If the index is significantly less than 1, the accepted model is the binomial distribution.  With both the Poisson and the binomial the amount of variance around the mean is no more than that which can be expected from random error.  If the index is significantly greater than 1, the distribution is considered contagious, where the variance in excess of that arising from random error is due to the probabilistic heterogeneity of the elements in the set, a success-breeding-success mechanism such as preferential attachment, or a combination of these two stochastic processes, i.e., the stochastic processes operative in the negative binomial and power-law distributions.  Table 2 below shows that the index of dispersion for Krugman's h-index truncated distribution is 1,813.47, placing it definitely in the contagious category.

Table 1. Relationship of full distributions to h-index truncated distributions of Google Scholar (GS) citations to the works of economist laureates

| Prize Winner | | | 1st 4% Bin Interval | Start of Asymptote | Number Works | | | Citation Range | | | Total Citations | | |
|---|---|---|---|---|---|---|---|---|---|---|---|---|---|
| Name | Year | H-Index | | | Full | H-Truncated | % H-Truncated | Full | H-Truncated | % H-Truncated | Full | H-Truncated | % H-Truncated |
| **Original Economist Sample Downloaded September 2011** | | | | | | | | | | | | | |
| Harry Markowitz | 1990 | 29 | 0 - 569 | 1st 4% Bin | 255 | 29 | 11.40% | 14219 | 14190 | 99.80% | 21157 | 20071 | 94.90% |
| James Heckman | 2000 | 107 | 0 - 517 | 1st 4% Bin | 941 | 107 | 11.40% | 12910 | 12801 | 99.20% | 70898 | 59281 | 83.60% |
| Paul Krugman | 2008 | 114 | 2 - 291 | 1st 4% Bin | 950 | 114 | 12.00% | 7220 | 7109 | 98.50% | 93827 | 80546 | 85.80% |
| Elinor Ostrom | 2009 | 77 | 0 - 475 | 1st 4% Bin | 912 | 77 | 8.40% | 11869 | 11791 | 99.30% | 43528 | 35813 | 82.30% |
| Peter Diamond | 2010 | 61 | 0 - 112 | 1st 4% Bin | 501 | 61 | 12.20% | 2793 | 2726 | 97.60% | 21828 | 19118 | 87.60% |
| **Validating Economist Sample Downloaded October 2013** | | | | | | | | | | | | | |
| Robert William Fogel | 1993 | 35 | 0-47 | 1st 4% Bin | 467 | 35 | 7.49% | 1162 | 1128 | 97.07% | 9755 | 7982 | 81.82% |
| C. W. J. Granger | 2003 | 82 | 0-851 | 1st 4% Bin | 391 | 82 | 20.97% | 21265 | 21180 | 99.60% | 79033 | 74793 | 94.64% |
| Thomas J. Sargent | 2011 | 80 | 0-136 | 1st 4% Bin | 870 | 80 | 9.20% | 3377 | 3296 | 97.60% | 38006 | 30914 | 81.34% |
| Lloyd S.Shapley | 2012 | 45 | 0-189 | 1st 4% Bin | 207 | 45 | 21.74% | 4712 | 4668 | 99.07% | 22957 | 21926 | 95.51% |
| Eugene F. Fama | 2013 | 83 | 0-471 | 1st 4% Bin | 382 | 83 | 21.73% | 11774 | 11691 | 99.30% | 134894 | 132647 | 98.33% |

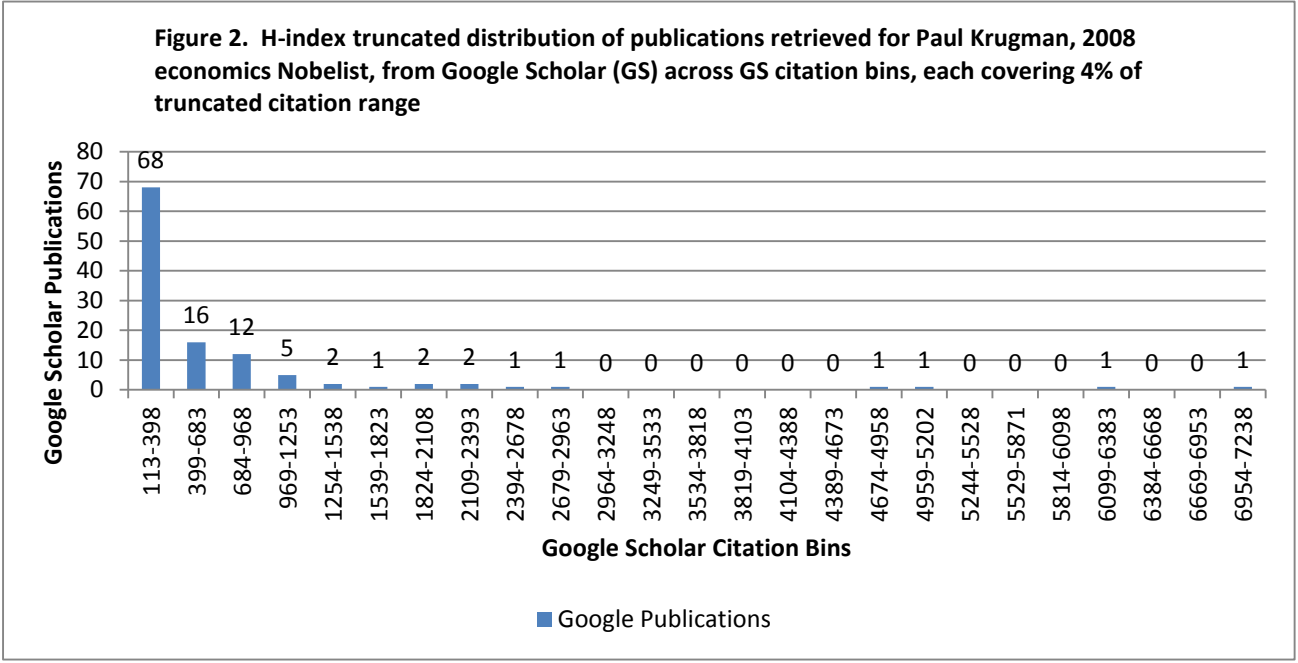

| Table 2. Analysis of the h-index distributions of Google Scholar (GS) citations to works of the winners of the Nobel Prize in economics | | | | | | | | | | | |
|---|---|---|---|---|---|---|---|---|---|---|---|
| Prize Winner | | | | Citation Range | | | | Citation Distributional Characteristics* | | | |
| Name | Year | H-Index | No. Works | Minimum | Maximum | Total Range | Total Citations | Mean | Variance | Index of Dispersion | Type |
| Original Economist Sample Downloaded September 2011 | | | | | | | | | | | |
| Harry Markowitz | 1990 | 29 | 29 | 29 | 14218 | 14190 | 20071 | 692.1 | 6824849.17 | 9861.02 | Contagious |
| James Heckman | 2000 | 107 | 107 | 109 | 12909 | 12801 | 59281 | 554.03 | 1664528.82 | 3004.41 | Contagious |
| Paul Krugman | 2008 | 114 | 114 | 114 | 7222 | 7109 | 80456 | 705.75 | 1279863.8 | 1813.47 | Contagious |
| Elinor Ostrom | 2009 | 77 | 77 | 78 | 11868 | 11791 | 35813 | 465.1 | 1856812.73 | 3992.25 | Contagious |
| Peter Diamond | 2010 | 61 | 61 | 67 | 2792 | 2726 | 19118 | 313.41 | 188994.68 | 603.03 | Contagious |
| Validating Economist Sample Downloaded October 2013 | | | | | | | | | | | |
| Robert William Fogel | 1993 | 35 | 35 | 34 | 1162 | 1128 | 20071 | 228.06 | 91307.41 | 400.37 | Contagious |
| C. W. J. Granger | 2003 | 82 | 82 | 85 | 21265 | 21180 | 59281 | 912.11 | 6936341.53 | 7604.72 | Contagious |
| Thomas J. Sargent | 2011 | 80 | 80 | 81 | 3377 | 3296 | 80456 | 386.43 | 268401.56 | 694.58 | Contagious |
| Lloyd S. Shapley | 2012 | 45 | 45 | 44 | 4712 | 4668 | 35813 | 487.24 | 745052.33 | 1529.11 | Contagious |
| Eugene F. Fama | 2013 | 83 | 83 | 83 | 11774 | 11691 | 19118 | 1598.16 | 7482706.04 | 4682.09 | Contagious |

* The distributions were tested with the chi-squared index of dispersion test for the Poisson to determine whether they were random or contagious. A random distribution would be either the binomial or Poisson, whereas a contagious distribution would be of the compound Poisson type like the negative binomial or Yule-Simon resulting from probabilistic heterogeneity and/or contagion. The test is based upon the Poisson identity that the lambda, mean, and variance equal each other. To do the test, the index of dispersion is calculated by dividing the theoretical variance represented by the mean into the actual variance. If the result is one, then the distribution is the Poisson; if the result is significantly below one, then distribution is binomial; and, if the result is significantly above one, then the distribution is considered contagious.

| Table 3. Test of fit of h-truncated distributions of Google Scholar (GS) citations to works of economist laureates to power-law distribution using MatLab software developed by Aaron Clauset*: $x_{min}$ = h-index ||||  |
|---|---|---|---|---|
| Prize Winner || Lotka Exponent | Kolmogorov-Smirnov p-value** | Fit to Power Law |
| Name | Year ||||
| **Original Economist Sample Downloaded September 2011** |||||
| Harry Markowitz | 1990 | 1.66 | 0.624 | Fit |
| James Heckman | 2000 | 1.91 | 0.007 | Rejected |
| Paul Krugman | 2008 | 1.82 | 0.108 | Possible |
| Elinor Ostrom | 2009 | 1.93 | 0.157 | Possible |
| Peter Diamond | 2010 | 1.92 | 0.03 | Most Likely Rejected |
| **Validating Economist Sample Downloaded October 2013** |||||
| Robert William Fogel | 1993 | 1.79 | 0.128 | Possible |
| C. W. J. Granger | 2003 | 1.72 | 0.178 | Possible |
| Thomas J. Sargent | 2011 | 1.93 | 0.289 | Possible |
| Lloyd S. Shapley | 2012 | 1.87 | 0.468 | Possible |
| Eugene F. Fama | 2013 | 1.93 | 0.242 | Possible |

*The Clauset software can be downloaded from: http://tuvalu.santafe.edu/~aaronc/powerlaws/. It is the basis of the analysis in the seminal article on power-law fitting: Clauset, A., Shalizi, C. R., & Newman, M. E. J. (2009). Power-law distributions in empirical data. SIAM Review, 51 (4), 661-703.

**A p-value below 0.05 indicates rejection of power-law hypothesis. A p-value above 0.05 indicates that the power-law hypothesis cannot be rejected.

Table 3 above gives the results of power-law tests of the h-index truncated distributions, using the MatLab software developed Clauset with the h-index equaling $x_{min}$. Krugman's estimated Lotkaian exponent was 1.82—definitely within the power-law range—but the Kolmogorov-Smirnov p-value indicates his h-index truncated distribution fitting a power-law is only possible but not definite. This corroborates Krugman's own experience with power-law distributions, and the findings of Clauset, Shalizi, and Newman (2009) indicate that the lack of precise fit may be due to the "messy" extended tail of his distribution.

The final two distributional tests were to estimate whether Krugman's h-index truncated GS citations distribution lies within the Gaussian or Lotkaian domain. To do this, his GS citations were

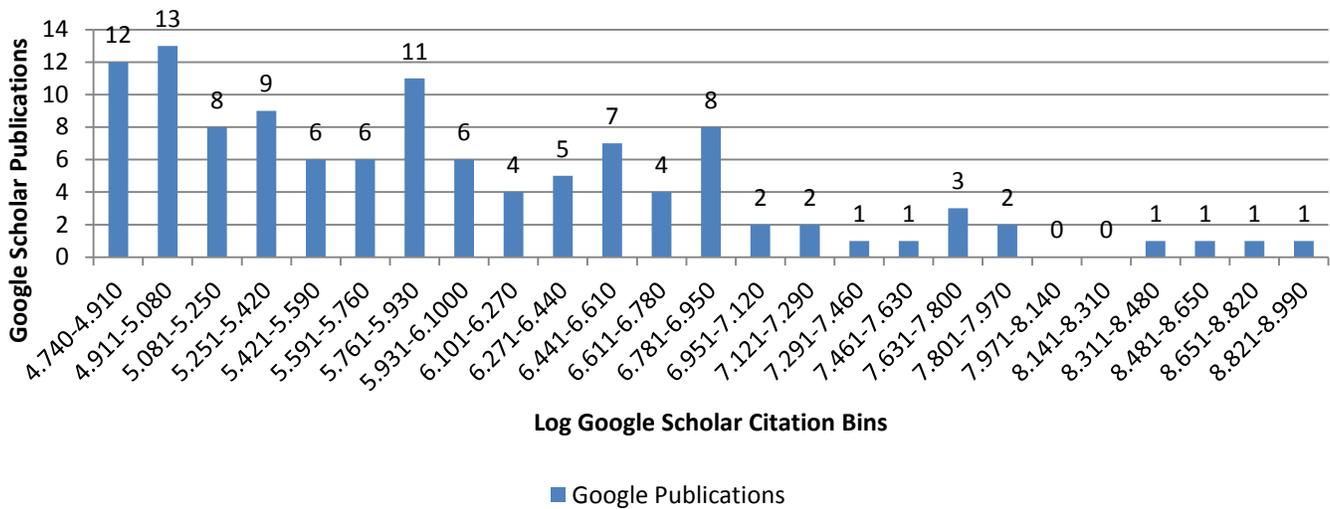

Figure 3. Frequency distribution of h-index truncated publications retrieved for Paul Krugman, 2008 economics Nobelist, from Google Scholar (GS) across log Google Scholar citation bins, each occupying 4% of the log citation range

subjected to the log + 1 transformation, and two histograms were constructed by dividing the logarithmic range into 25 equal 4% bins. As stated above, with the first histogram, only the X-axis was on the logarithmic scale. The Y-axis scale was not changed to the logarithmic scale, so that histogram would graph the frequency counts of his work over the logarithmic bins. The resulting histogram is shown in Figure 3. The logarithmic binning of the data established control over the messy tail for antilog calculations showed that the bins exponentially increase in width from 16 for the 1st 4% logarithmic bin to 1,239 for the 25th 4% logarithmic bin, consolidating the four outliers on the right by placing them in neighboring bins and bringing these bins leftward closer to the main body of observations. Figure 3 above shows that the logarithmic histogram does not approximate the bell-shaped curve of the normal distribution but maintains the same negative slope as the untransformed distribution. This is evidence that the distribution does not conform to the lognormal and is not within the Gaussian domain. Table 4 below sets forth results of the Shapiro-Wilks tests of whether the logarithmically transformed data fits

| Table 4. Shapiro-Wilks test of fit of logarithmically transformed distributions of Google Scholar (GS) citations to works of economist laureates to the normal distribution ||||
|---|---|---|---|
| **Prize Winner** || **Shapiro-Wilks p-value*** | **Fit to Normal Distribution** |
| **Name** | **Year** | | |
| **Original Economist Sample Downloaded September 2011** ||||
| Harry Markowitz | 1990 | 0.001 | Rejected |
| James Heckman | 2000 | < 0.001 | Rejected |
| Paul Krugman | 2008 | < 0.001 | Rejected |
| Elinor Ostrom | 2009 | < 0.001 | Rejected |
| Peter Diamond | 2010 | < 0.001 | Rejected |
| **Validating Economist Sample Downloaded October 2013** ||||
| Robert William Fogel | 1993 | 0.005 | Rejected |
| C. W. J. Granger | 2003 | < 0.001 | Rejected |
| Thomas J. Sargent | 2011 | < 0.001 | Rejected |
| Lloyd S. Shapley | 2012 | 0.002 | Rejected |
| Eugene F. Fama | 2013 | 0.001 | Rejected |
| *A Shapiro-Wilks p-value below 0.01 designates a rejection of the normal distribution at the 1% level. ||||

fits the normal distribution. This standard test for normality showed that Krugman's logarithmically transformed h-index truncated GS citations distribution had a <0.001 probability of fitting the normal distribution, refuting the lognormal alternative hypothesis and strengthening the case for his distribution being within the Lotkaian domain.

To test whether Krugman's h-index truncated GS citation distribution was within the Lotkaian domain, the scale of the Y-axis of the histogram was changed to the logarithmic scale, and the logged number of citations on the X-axis was regressed on the logged number of works on the Y-axis. The method being utilized here has been theoretically justified by Milojević (2010b), who advocates partial logarithmic binning as the method of choice for uncovering the nature of many distributions encountered in information science. According to her, logarithmic binning retrieves information and

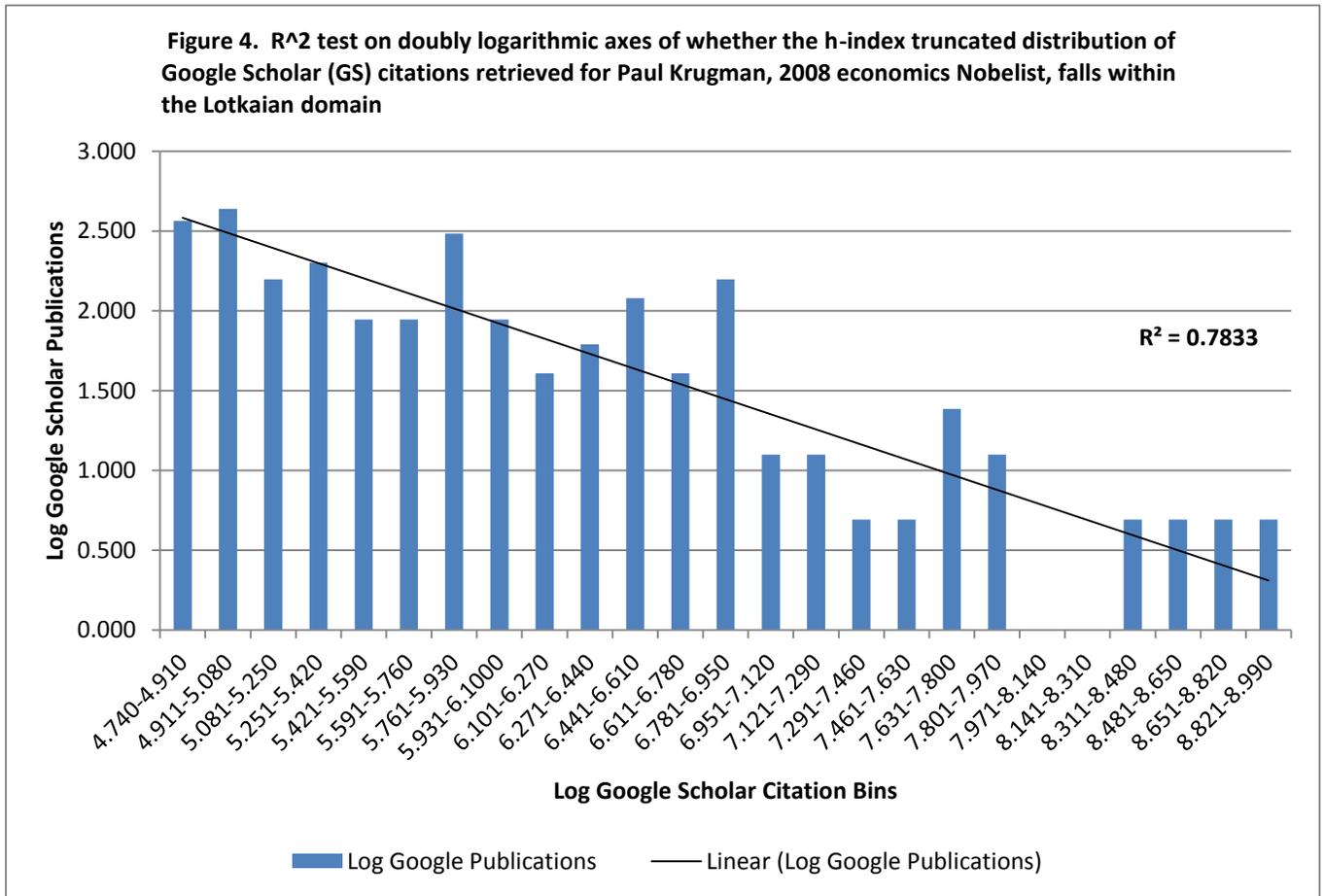

Figure 4. R^2 test on doubly logarithmic axes of whether the h-index truncated distribution of Google Scholar (GS) citations retrieved for Paul Krugman, 2008 economics Nobelist, falls within the Lotkaian domain

F=83.13. Fit is significant at the 1% level
Slope/Lotka Exponent = -1.41

trends invisible in messy noisy power law tails. According to her, obtaining the exponent from logarithmically binned data using a simple least square method is in some cases warranted in addition to methods such as the maximum likelihood. Most interestingly, she stipulates that the least squares method should not be applied to un-binned data and that the distribution should be truncated on the left as we are doing by setting the h-index at $x_{min}$. She concludes that binning also allows an unbiased power law exponent to be determined using the traditional least square method and without discarding data in the tail, and she recommends obtaining the exponent from the binned data in addition to procedures such as the maximum likelihood method, which is preferable when the underlying distribution is known to follow a pure power law—something that is never known and rarely proven.

| Prize Winner | Year | Slope/Lotka Exponent* | R^2** | F*** |
|---|---|---|---|---|
| **Table 5. Estimates of the Lotka exponent and linear fits (R^2) to the trendline resulting from regressing log Google Scholar (GS) citations of Nobelists in economics on their log GS publications for distributions truncated at the h-index** | | | | |
| **Original Economist Sample Downloaded September 2011** | | | | |
| Harry Markowitz | 1990 | -1.85 | 0.4106 | 16.02 |
| James Heckman | 2000 | -1.32 | 0.8519 | 132.23 |
| Paul Krugman | 2008 | -1.41 | 0.7833 | 83.13 |
| Elinor Ostrom | 2009 | -1.51 | 0.8217 | 106.02 |
| Peter Diamond | 2010 | -1.10 | 0.5061 | 23.56 |
| **Validating Economist Sample Downloaded October 2013** | | | | |
| Robert William Fogel | 1993 | -0.841 | 0.2010 | 5.784 |
| C. W. J. Granger | 2003 | -1.72 | 0.7769 | 80.108 |
| Thomas J. Sargent | 2011 | -1.14 | 0.6415 | 41.16 |
| Lloyd S. Shapley | 2012 | -1.48 | 0.4494 | 18.77 |
| Eugene F. Fama | 2013 | -1.27 | 0.2855 | 9.19 |

*In power-law distributions such as that of Lotka the exponent equals the absolute value of the negative slope of the regression or trendline in a log-log plot.

** The Excel LINEST function uses regression analysis to calculate the linear trendline, using the "least squares" method to find the straight line that best fits the data. R^2--or the coefficient of determination--runs from 0 or no fit of the line to the data to 1 or a perfect fit of the line to the data.

*** For significance of the fit, LINEST not only produces the F statistic but also a degrees of freedom (df) statistic that enables calculating the v1 and v2 degrees of freedom necessary to enter published F-distribution tables at the proper place to determine the significance level. The formulas are v1 = n-df-1 and v2 = df, where n is the number of data points and df is the LINEST degrees of freedom. Our data was organized in such a way that n always equaled 25 and df always equaled 23. Therefore, v1 always equaled 1, and v2 always equaled 23. As a result, the same level of significance is applicable to all the tests above. To be significant, F has to be 4.28 and above at the 5% level or 7.88 and above at the 1% level.

The results of the test for Krugman are graphed in Figure 4 and summarized in Table 5 above. Figure 4 shows a regression line connecting the logarithmic citation bins with a negative slope or Lotka exponent of 1.41. This is lower than the maximum likelihood estimate of 1.82 but still within the power-limits set by Rousseau and Rousseau (2000). The crucial point is that the exponent is below the 3.0 indicating that it is within the Lotkaian domain and below the Gaussian domain. The $R^2$ fit to this line is 0.7833—fairly close to the ultimate limit of 1.0--that is significant at the 1% level. The evidence is that Krugman's h-index truncated GS citation distribution is within the Lotkaian domain.

| Table 6. Bibliographic structure of h-index publications of economist Nobelists | | | | | | | | | | | |
|---|---|---|---|---|---|---|---|---|---|---|---|
| Prize Winner | | | | Bibliographic Categories | | | | | | | |
| Name | Year | H-Index | Sample Size | Monographs | % Monographs | Journal Articles | % Journal Articles | Book Chapters | % Book Chapters | Working Papers | % Working Papers |
| **Original Economist Sample Downloaded September 2011** | | | | | | | | | | | |
| Harry Markowitz | 1990 | 29 | 15 | 3 | 20.0% | 8 | 53.3% | 0 | 0.0% | 4 | 26.7% |
| James Heckman | 2000 | 107 | 36 | 1 | 2.8% | 21 | 58.3% | 2 | 5.6% | 12 | 33.3% |
| Paul Krugman* | 2008 | 114 | 38 | 9 | 23.7% | 14 | 36.8% | 6 | 15.8% | 7 | 18.4% |
| Elinor Ostrom | 2009 | 77 | 26 | 6 | 23.1% | 11 | 42.3% | 7 | 26.9% | 2 | 7.7% |
| Peter Diamond | 2010 | 61 | 21 | 4 | 19.0% | 12 | 57.1% | 3 | 14.3% | 2 | 9.5% |
| **Validating Economist Sample Downloaded October 2013** | | | | | | | | | | | |
| Robert William Fogel | 1993 | 35 | 15 | 3 | 20.0% | 7 | 46.7% | 1 | 6.7% | 4 | 26.7% |
| C. W. J. Granger | 2003 | 82 | 28 | 4 | 14.3% | 22 | 78.6% | 1 | 3.6% | 1 | 3.6% |
| Thomas J. Sargent | 2011 | 80 | 24 | 3 | 12.5% | 17 | 70.8% | 0 | 0.0% | 4 | 16.7% |
| Lloyd S. Shapley | 2012 | 45 | 23 | 1 | 4.3% | 12 | 52.2% | 4 | 17.4% | 6 | 26.1% |
| Eugene F. Fama | 2013 | 83 | 28 | 0 | 0.0% | 28 | 100.0% | 0 | 0.0% | 0 | 0.0% |
| *Two--or 5.3%--of Krugman's h-index publications were New York Times op-ed pieces. | | | | | | | | | | | |

The statistical analyses of the h-index truncated distributions of the economist laureates were buttressed by investigations of the bibliographic and authorship structures of the works comprising their h-index. The sets for these investigations were formed by taking stratified random samples starting with the work highest in GS citations of every third work for laureates with high h-indexes and every second work for laureates with low h-indexes. Table 6 above sets forth the results of the investigation of the bibliographic structure. The results for Krugman shown in the table are as follows: 9 or 23.7% of his h-index works were monographs; 14 or 36.8% were journal articles; 6 or 15.% were book chapters; and 7 or 18.4 % were working papers. A working paper here is defined as a preliminary scientific or technical paper that authors release to share ideas about a topic or to elicit feedback that is often published by a think tank such as the National Bureau of Economic Research, American Enterprise Institute, Rand Corporation, etc. Two or 5.3% of Krugman's h-index works were *New York Times* op-ed pieces.

| Table 7. Authorship structure of h-index publication of economist Nobelists ||||||||||||
| Prize Winner || | | Number Authors || Authorship Position ||||||
| Name | Year | H-Index | Sample Size | Range of Number of Authors | Median Number of Authors | Sole | % Sole | Primary | % Primary | Secondary | % Secondary |
| --- | --- | --- | --- | --- | --- | --- | --- | --- | --- | --- | --- |
| Original Economist Sample Downloaded September 2011 ||||||||||||
| Harry Markowitz | 1990 | 29 | 15 | 1 to 4 | 2.5 | 7 | 46.7% | 3 | 20.0% | 5 | 33.3% |
| James Heckman | 2000 | 107 | 36 | 1 to 4 | 2 | 9 | 25.0% | 18 | 50.0% | 9 | 25.0% |
| Paul Krugman | 2008 | 114 | 38 | 1 to 4 | 1 | 29 | 76.3% | 5 | 13.2% | 4 | 10.5% |
| Elinor Ostrom | 2009 | 77 | 26 | 1 to 4 | 2 | 10 | 38.5% | 0 | 0.0% | 16 | 61.5% |
| Peter Diamond | 2010 | 61 | 21 | 1 to 4 | 2 | 10 | 47.6% | 6 | 28.6% | 5 | 23.8% |
| Validating Economist Sample Downloaded October 2013 ||||||||||||
| Robert William Fogel | 1993 | 35 | 15 | 1 to 10 | 1 | 14 | 93.3% | 1 | 6.7% | 0 | 0.0% |
| C. W. J. Granger | 2003 | 82 | 28 | 1 to 5 | 2 | 8 | 28.6% | 10 | 35.7% | 10 | 35.7% |
| Thomas J. Sargent | 2011 | 80 | 24 | 1 to 4 | 2 | 9 | 37.5% | 3 | 12.5% | 12 | 50.0% |
| Lloyd S. Shapley | 2012 | 45 | 23 | 1 to 3 | 2 | 7 | 30.4% | 6 | 26.1% | 10 | 43.5% |
| Eugene F. Fama | 2013 | 83 | 28 | 1 to 3 | 2 | 12 | 42.9% | 13 | 46.4% | 3 | 10.7% |

The results of the investigation of authorship structure of the laureates' h-index works are given in Table 7 above. Here it is seen that: 1) Krugman's h-index works had a range of 1 to 4 authors with a median of 1 author: Krugman was sole author 29 times (76.3%); primary author 5 times (13.2%); and he was secondary author 4 times (10.5%).

| Table 8. Outlier works of original economist sample downloaded September, 2011, at extreme right tip of Google Scholar (GS) cites tail | | | | |
|---|---|---|---|---|
| Work | Type of Work | Authorship Positon | GS Cites | % Total H-Index Cites |
| **Harry Markowitz 1990** | | | | |
| *Portfolio selection: efficient diversification of investments* / Harry Markowitz. New York: Wiley, 1959. | Monograph | Sole | 14218 | 70.8% |
| **James Heckman 2000** | | | | |
| Sample Selection Bias as a Specification Error / James Heckman. *Econometrica*, Vol. 47, No. 1 (Jan., 1979), pp. 153-161. | Journal Article | Sole | 12909 | 21.8% |
| **Paul Krugman 2008** | | | | |
| *Geography and trade* / Paul Krugman. Cambridge, Mass.: MIT Press, 1991. | Monograph | Sole | 7222 | 9.0% |
| *Increasing returns and economic geography* / Paul Krugman. Cambridge, Mass.: National Bureau of Economic Research, 1990. NBER working paper series; working paper no. 3275. | Working Paper | Sole | 6242 | 7.7% |
| *The spatial economy: cities, regions and international trade* / Masahisa Fujita, Paul Krugman, Anthony J. Venables. Cambridge, Mass.: MIT Press, 1999. | Monograph | 2nd of 3 | 5072 | 6.3% |
| *Market structure and foreign trade: increasing returns, imperfect competition, and the international economy* / Elhanan Helpman and Paul R. Krugman. Cambridge, Mass.: MIT Press, 1985. | Monograph | 2nd of 2 | 4749 | 5.9% |
| | | Total = | 23285 | 28.9% |
| **Elinor Ostrom 2009** | | | | |
| *Governing the commons: the evolution of institutions for collective action* / Elinor Ostrom. Cambridge, England: Cambridge University Press, 1990. | Monograph | Sole | 11868 | 33.1% |
| **Peter Diamond 2010** | | | | |
| National Debt in a Neoclassical Growth Model / Peter Diamond. *The American Economic Review*, Vol. 55, No. 5, Part 1 (Dec., 1965), pp. 1126-1150. | Journal Article | Sole | 2792 | 14.6% |

A signature feature of the h-index truncated distributions of the economist laureates are the extreme outliers at the right tip of the asymptote that seem to encapsulate the essence of their contribution to the discipline. It is this "messy tail" that Clauset, Shalizi, and Newman (2009) posit as a primary reason for the difficulty in obtaining proper fits to power-law distributions. Therefore, the composition of these outliers is of primary interest, and it is set forth for the original 2011 sample of economist in Table 8 above. Krugman had four such outliers, and their characteristics are summarized in Table 8. These outliers account for a total of 23,285 GS cites or 28.9% of his h-index truncated GS cites. None of these works are *New York Times* op-ed pieces, for which he is more widely known, and all are academic works dealing with topics for which he was awarded the Nobel prize. Three of these are monographs published by MIT Press, and one is a crudely printed National Bureau of Economic Research working paper, confirming his negative assessment of the role of highly ranked journals in the advancement of his discipline. Krugman is sole author twice, second of two authors once, and second of three once, confirming GS's ability to retrieve a researcher's work no matter his authorship position and raising the question of allocation of credit. These results are certainly not random and confirm GS's ability to form relevant sets—particularly the fact that the outliers are on the topic for which he was awarded the prize and do pinpoint his contribution to the discipline.

**Data Analysis: Comparison of the Original 2011 Laureate Sample with the 2013 Validating Sample**

In this section we will compare the original 2011 sample of economist laureates with the validating 2013 sample of economists to determine whether all laureates follow the same basic model as Krugman. However, before this can be done, it is necessary to point out that two of the laureates have h-indexes substantially below 50—the sample size of works required by Clauset, Shalizi, and Newman (2009) for reliable estimates of parameters. The first is Markowitz (1990) in the original 2011 laureate sample, whose h-index is only 29. The second is Fogel (1993) in the validating 2013 laureate sample. When his data was downloaded, his h-index was 46, but so much error was found in his data that editing and cleansing it reduced his h-index to 35. As a result of their low h-indexes, their results can be expected to be somewhat erratic. It is interesting to note that both Markowitz and Fogel are the oldest in terms of prize year in their samples and by far the oldest of all the other prize winners in these terms. Given the growing importance of GS citations in evaluating researchers, what is of most interest here is that most researchers have GS citation h-indexes far below 50, and, if Markowitz and Fogel follow the same general model as Krugman, it could be considered a validation of their evaluations.

The histograms modeling the full GS citation distributions of the other nine economist Nobelists came out be basically the same as the one that resulted for Krugman and is shown in Figure 1. Here the primary two features are: 1) the heavy concentration of works in the 1$^{st}$ or lowest bin covering 4% of the total citation range; and 2) the long, exponentially increasing asymptote or "tail" to the right. An analysis of the data on the full distribution set forth in Table 1 reveals that both the other laureates in the 2011 original sample and all the laureates in the 2013 validating sample followed the same basic pattern as Krugman. Thus, for all the other laureates in both samples, the h-index occurs in the first bin demarcating the lowest 4% of GS citations range, where the asymptote or "tail" also originates. Setting the x$_{min}$ at the h-index to demarcate the h-index truncated range or the "tail," the figures for the other laureates in the original 2011 sample are as follows: Markowitz—the h-index truncated distribution contains 11.4% of the total works but covers 99.8% of the total citation range and has 94.9% of the total citations; Heckman--the h-index truncated distribution contains 11.4% of the total works but covers 99.2% of the total citation range and has 83.6% of the total citations; Ostrom--the h-index truncated distribution contains 8.4% of the total works but covers 99.3% of the total citation range and has 82.3% of the total citations; and Diamond—the h-index truncated distribution contains 12.2% of the total works but covers 97.6% of the total citation range and has 87.6% of the total citations. The figures for the laureates in the 2013 validating sample of laureates are as follows: Fogel—the h-index truncated distribution contains 7.5% of the total works but covers 97.1% of the total citation range and has 81.8% of the total citations; Granger—the h-index truncated distribution contains 21.0% of the total works but covers 99.6% of the total citation range and has 94.6% of the total citations; Sargent—the h-index truncated distribution contains 9.2% of the total works but covers 97.6% of the total citation range and has 81.3% of the total citations; Shapley—the h-index truncated distribution contains 21.7% of the total works but covers 99.1% of the total citation range and has 95.5% of the total citations; and Fama—the h-index truncated distribution contains 21.7% of the total works but covers 99.3% of the total citation range and has 98.3% of the total citations. The data makes abundantly clear that the sets of works demarcating both the h-index and the "tail" are conterminous, and, if one takes into account that the so-called "works" below the h-index are increasingly irrelevant and full of error as their citations decrease, this serves as proof not only of the h-index's validity but also the ability of Google Scholar to define relevant documentary sets.

     Similarly the histograms of the h-index truncated distributions of the other 9 laureates came out to be basically the same as the one shown in Figure 2 for Krugman--modeling an extremely right-skewed, reversed-J-shaped or negative exponential distribution. Like Krugman's, they all had what appears to be a signature feature of h-index truncated economist GS citation distributions—extremely "messy" tails due to a number extreme outliers at the right tip of the asymptote. Table 2 sets forth the statistics on their h-index truncated GS citation distributions. Here the key feature is that the chi-squared index of dispersion test revealed that all these distributions were highly "contagious" or resulting from the same stochastic processes of probabilistic heterogeneity and/or "contagion"—i.e., success-breeds-success, preferential attachment, etc.—that is operative in the negative binomial, power-law, and other types of compound-Poisson distributions. This makes of great interest Table 3, which summarizes the test results for the power-law fits of the laureates' h-index truncated GS citation distributions, using the MatLab software developed by Aaron Clauset with $x_{min}$ set to equal the h-index. As can be seen, the Lotka exponents of all the Nobelists—both of the original 2011 and validating 2013

samples—were estimated to range from 1.66 to 1.93, i.e., below the critical 3.0 and in the Lotkaian or power-law domain. However, there were notable differences between the original 2011 and validating 2013 samples in the Kolmogorov-Smirnov estimates of their power-law fits. In the first sample, Markowitz with a p of 0.624 appears to be a "fit," but his h-index is only 29—far below the required sample size of 50, making his results possibly erratic. Therefore, Markowitz at best should be considered only a "possible" fit. Diamond's p of 0.03 is close enough to the required 0.05 to be considered a "possible" fit, but, to be conservative, we will classify him as "rejected" together with Heckman. Thus for the original 2011 sample, we have three "possible" and two "rejected." There were no such problems with the 2013 validating sample, and all their h-index truncated GS citation distributions tested to be "possible" fits to a power-law. It will be seen below that this was a questionable result for Fogel, whose h-index of 35 is far below the required sample of 50.

    The somewhat ambiguous results of the Kolmogorov-Smirnov tests for fits to power-law distributions or the "Lotkaian domain" make of extreme interest the tests for the most commonly proposed alternative hypothesis—lognormal fits or the "Gaussian domain." As stated above, to conduct these tests, the GS citations of all the laureates of both the 2011 original and 2013 validating samples were logarithmically transformed. All the histograms graphically showing the results of this operations came out to be basically the same as the one done for Krugman shown in Figure 3—no sign of the signature bell-shaped curve and greater control over the "messy" outliers at the right tip of the "tail." Table 4 shows that fits to the normal law of error were solidly rejected for the h-index truncated GS citation distributions of all the laureates by the standard Shapiro-Wilks test for normality. This unanimous rejection of the "Gaussian domain" strengthens the case for the "Lokaian domain."

    With one glaring exception, utilization of the traditional method of identifying a power-law distribution by converting both axes to the logarithmic scale, regressing the items on the X-axis on the sources on the Y-axis, and then obtaining the $R^2$ fit of the data to the resulting regression line

corroborated the above favorable but ambiguous results in favor of the laureates' h-index truncated GS citation distributions being within the Lotka domain. All the histograms constructed with this procedure for the laureates basically replicated—but with key differences—the one constructed for Krugman and shown in Figure 4. Table 5 summarizes the statistical results of the traditional method. Here it can be seen that the estimated Lotka exponents were generally lower than those derived with Clauset's MatLab techniques, but—with the one glaring exception—all were above or close to the lower limit of 1.26 set by Rousseau and Rousseau (2000) as the lower limit for such exponents. The one glaring exception was Fogel, whose exponent was 0.841, but it was noted above that Fogel's h-index of the 35 was much lower than the required sample n of 50 and that his results could be expected to be erratic. The same may hold true for Markowitz, who has the highest exponent of 1.85 but an h-index of merely 29. The important thing here is that all the exponents or slopes are negative, corroborating the above tests that rejected the alternative hypothesis of the Gaussian domain. The key differences in the histograms were the widely differing $R^2$ fits to the regression lines, which ranged from a low of 0.2010 for Fogel to a high of 0.8127 for Ostrom. In a coming paper we will show that these $R^2$ differences are not meaningless but—given the probable rarity of close $R^2$ fits to power-law distributions that require enough time and proper conditions—a crude measure of the extent to which a researcher's ideas have been incorporated into his discipline.

In respect to the bibliographic and authorship structure of the laureates' h-index works Tables 6 and 7 show that these are basically the same for both the 2011 original and 2013 validating samples. Concerning the bibliographic structure, Table 6 shows that journal articles were the most important, ranging from a low of 36.8% for Krugman (2008) to a high of 100.0% for Fama (2013). As can be expected of a social science, monographs were also important, ranging from a low of 0.0% for Fama (2013) to a high of 23.1% for Ostrom (2009). However, what is most significant in Table 6 is that it shows that working papers were more important than monographs, ranging from 0.0% for Fama (2013) to a high of 33.3% for Heckman. This corroborates Krugman's observations on the importance of this bibliographic medium—and by extension today's blogs—in the dissemination of new economic ideas. As for authorship structure, Table 7 shows that number of authors was about the same for both samples with a median of around 2. It also shows that the laureates were mostly either sole or primary authors, except for Sargent (2011), who was secondary author 50% of the time, and Ostrom (2009), who was secondary author 61.5% of the time. It is of interest to note here that Ostrom was the only female laureate.

A signature feature of the h-index truncated GS distributions of the economist laureates are the extreme outliers at the right tip of the asymptote, causing the "tail" to be "messy" and preventing a precise fit of these distributions to the Lotkaian or power-law model. Therefore, it is of interest to determine whether these works are related to the reasons for which the laureates were awarded their prize. If so, it would be conclusive evidence that GS document sets are relevant. The reasons why the original 2011 sample of laureates are set forth below (Nobelprize.org, 2014):

**Markowitz (1990)**: **Field**: financial economics; **Prize motivation**: for pioneering work in the theory of financial economics; **Contribution**: Constructed a micro theory of portfolio management for individual wealth holders.

**Heckman (2000)**: **Field**: econometrics; **Prize motivation**: for his development of theory and methods for analyzing selective samples; **Contribution**: Developed methods for handling selective samples in a statistically satisfactory way.

**Krugman (2008)**: **Field**: international and regional economics; **Prize motivation**: for his analysis of trade patterns and location of economic activity; **Contribution**: Integrated the previously disparate research fields into a new, international trade and economic geography.

**Ostrom (2009): Field:** political scientist studying economic governance; **Prize motivation:** for her analysis of economic governance, especially the commons**; Contribution**: Challenged the conventional wisdom by demonstrating how local property can be successfully managed by local commons without any regulation by central authorities or privatization.

**Diamond (2010): Field**: labor economics; **Prize motivation**: for his analysis of markets with search frictions.

The extreme outliers for the original 2011 are bibliographically and statistically analyzed in Table 8 above, and, except for Diamond, these works are clearly related to the reasons the laureates were awarded the prize.

The reasons why the laureates of the validating 2013 sample were awarded the Nobel Prize are given below (Nobelprize.org, 2014):

**Fogel** (1993): **Field:** economic history;  **Prize motivation**: for having renewed research in economic history by applying economic theory and quantitative methods in order to explain economic and institutional change; **Contribution**: Clarified the role of the railways for the development of the economy in the United States, and the economic role of slavery.

**Granger (2003): Field:** econometrics**; Prize motivation:** for methods of analyzing economic time series with common trends (cointegration)**; Contribution:** Developed and applied new statistical methods, based on so-called "cointegration", to differentiate between, and combine the analysis of, short-term fluctuations and long-term trends.

**Sargent (2011)**: **Field**: macroeconometrics; **Prize motivation**: for empirical research on cause and effect in the macroeconomy.

**Shapley (2012): Field:** game theory**; Prize motivation:** for the theory of stable allocations and the practice of market design**; Contribution:** succeeded in identifying methods of solving key economic problem of bringing different players into stable relationships.

| Table 9. Outlier works of validating economist sample downloaded October, 2013, at extreme right tip of Google Scholar (GS) cites tail | | | | |
|---|---|---|---|---|
| Work | Type of Work | Authorship Positon | GS Cites | % Total H-Index Cites |
| **Robert William Fogel 1993** | | | | |
| *Economic growth, population theory, and physiology: the bearing of long-term processes on the making of economic policy* / Robert William Fogel. Cambridge, Mass. : National Bureau of Economic Research, 1994 NBER working paper series; working paper no. 4638. | Working Paper | Sole | 1162 | 14.56% |
| *Time on the cross: the economics of American Negro slavery* / Robert William Fogel and Stanley L. Engerman. Boston: Little, Brown, 1974. | Monograph | 1st of 2 | 1095 | 13.72% |
| *Railroads and American economic growth: essays in econometric history* / Robert William Fogel. Baltimore: Johns Hopkins Press, 1964. | Monograph | Sole | 1028 | 12.88% |
| | | **Total =** | 3285 | 41.16% |
| **C. W. J. Granger 2003** | | | | |
| Co-integration and error correction: representation, estimation, and testing / Robert F. Engle and C. W. J. Granger. *Econometrica*, Vol. 55, No. 2 (Mar., 1987), pp. 251-276. | Journal Article | 2nd of 2 | 21265 | 28.43% |

| Table 9 [Continued]. Outlier works of validating economist sample downloaded October, 2013, at extreme right tip of Google Scholar (GS) cites tail | | | | |
|---|---|---|---|---|
| Work | Type of Work | Authorship Positon | GS Cites | % Total H-Index Cites |
| **Thomas J. Sargent 2011** | | | | |
| *Bounded rationality in macroeconomics./* Thomas J. Sargent. Oxford [England] : Clarendon Press, 1993. | Monograph | Sole | 3377 | 10.92% |
| **Lloyd S. Shapley 2012** | | | | |
| *A value for n-person games* / Lloyd S. Shapley. Santa Monica, California: The Rand Corporation, 1952. | Working Paper | Sole | 4712 | 21.49% |
| **Eugene F. Fama 2013** | | | | |
| Efficient capital markets: a review of theory and practice / Eugene F. Famam. *Journal of Finance*, Vol.25, Issue 2 (May, 1970), pp. 383-417.. | Journal Article | Sole | 11774 | 8.88% |
| Common risk factors in the returns on stocks and bonds / Eugene F. Fama and Kenneth R. French. *Journal of Financial Economics*, Vol. 33, Issue 1 (Feb., 1993), pp. 3-66. | Journal Article | 1st of 2 | 11721 | 8.84% |
| Separation of Ownership and Control / Eugene F. Fama and Michael C. Jensen. *Journal of Law and Economics*, Vol. 26, No. 2, Corporations and Private Property: A Conference Sponsored by the Hoover Institution (June, 1983), pp. 301-325 | Journal Article | 1st of 2 | 11488 | 8.66% |
| The cross-section of expected stock returns / Eugene F. Fama and Kenneth R. French. *The Journal of Finance*,Vol. 47, Issue 2 (Jun, 1992), pp. 427-465. | Journal Article | 1st of 2 | 10454 | 7.88% |
| | | Total = | 45437 | 34.25% |

**Fama** (2013): **Field**: financial economics; **Prize motivation**: for empirical analysis of asset prices; Contribution: demonstrated that stock price movements are impossible to predict in the short-term and that new information affects prices almost immediately, which means that the market is efficient, laying the basis for index funds.

The extreme outliers of the 2013 validating sample of laureates are bibliographically and statistically analyzed in Table 9 above, and it is quite obvious that they are all clearly related to reasons why these economists were granted the Nobel Prize. The relevancy of the GS document sets is here confirmed again.

**Conclusions**

This paper has presented an analysis of whether Google Scholar (GS) can construct documentary sets relevant for the evaluation of the works of researchers. It conducted this analysis in terms of total GS citations to these works. The researchers analyzed were two samples of Nobelists in economics: an original sample of five laureates downloaded in September, 2011; and a validating sample of laureates downloaded in October, 2013. Two methods were utilized to conduct this analysis. The first can be called distributional, and it consisted of discovering the type of the distribution governing the distributions of the laureates' works by GS citations and how did a key measure of quality—the laureates' h-index—relate to this type of distribution. It was found that all the laureates' distributions were in the power-law or Lotkaian domain, a key feature of which was an exponential asymptote or "tail" to the right along the x-axis demarcating the items or GS citations. It also found that the laureates' h-index—which theoretically defines the important œuvre of a given researcher—is an excellent estimate of the start of this asymptote and that the sets of works comprising the asymptote and the h-index are conterminous. This overlap validates both the ability of GS to define sets relevant for evaluating researchers and the h-index as defining the core œuvre of a researcher.

The second method of analyzing the ability of GS to construct documentary sets relevant for the evaluation of researchers was semantic. A signature feature of the distributions of all the Nobelist economists were extreme outliers at the extreme right tip of the asymptote. These outliers make the tail "messy" and have been hypothesized to prevent precise fits to power-law or Lotkaian distributions. These outliers were semantically analyzed and were found not to be random but related by subject to the contributions to the discipline for which the laureates were awarded their prize. This serves as validation of the ability of GS to define semantically by links the subject relevance of the documentary sets.

These were the major findings of this paper, and here the two samples of economists validated each other. However, besides these, the paper made one other major discovery and raised three important issues. In respect to the first, there was revealed the important role of working papers in disseminating new economic knowledge, corroborating Krugman's observations on this matter. Working papers were found to be about as important as monographs. In respect to the issues, both Markowitz and Fogel had h-indexes far below the required n for a valid sample of 50. However, they seem to follow the same basic pattern—though erratically—as the others, raising the issue of whether such methods can be applied to other researchers—the vast majority—whose h-indexes are far below 50. It was also seen that Nobelists often collaborate and even are secondary authors, raising the issue of the allocation of credit for a given work. And, finally, there appears to be a misunderstanding of the true purpose of the $r^2$ test for a power-law distribution, and this will require another paper, due to the complexities of this issue.

Hirsch, J. E. (2005). An index to quantify an individual's scientific research output. *PNAS*, 102 (46), 16569-16572.

Hirsch, J. E. (2007). Does the h index have predictive power? *PNAS*, 104 (49), 19193-19198.

Hirsch, J. E. (2010). An index to quantify an individual's scientific research output that takes into account the effect of multiple coauthorship. *Scientometrics*, 85 (3), 741-754.

Huberman, B. A. (2001). *The laws of the Web: Patterns in the ecology of information*. Cambridge, Mass.: MIT Press.

Ijiri, Y., & Simon, H. A. (1977). Skew distributions and the sizes of business firms. Studies in mathematical and managerial economics. Amsterdam: North-Holland.

Jacsó, P. ( 2011). Google Scholar duped and deduped—the aura of robometrics." *Online Information Review*, 35 (1), 154-160.

Krugman, P. (1996a). Confronting the mystery of urban hierarchy. *Journal of the Japanese and International Economies*, 10 (Article No. 0023), 399-418.

Krugman, P. (1996b). *The Self-Organizing Economy.* Cambridge, Mass.: Blackwell.

Krugman, P. (2012, Jan. 17). Open science and the econoblogosphere. *New York Times*. Retrieved June 20, 2014 from http://krugman.blogs.nytimes.com/2012/01/17/open-science-and-the-econoblogosphere/?_php=true&_type=blogs&_php=true&_type=blogs&_r=1

Lotka, A. J. (1926). The frequency distribution of scientific productivity. *Journal of the Washington Academy of Sciences*, 16 (12), 317-322.

Keynes, J. M. (1921). *A treatise on probability*. London: Macmillan.

MathWorks Documentation Center. (2014) MatLab: Lognormal distribution. Retrieved April 2, 2014 from: http://www.mathworks.com/help/stats/lognormal-distribution.html?searchHighlight=lognormal+distribution

McAlister, D. (1879). The law of the geometric mean. *Proceedings of the Royal Society of London*, 29, 367-376.

Milojević, S. (2010a). Modes of collaboration in modern science: beyond power laws and preferential attachment. *Journal of the American Society for Information Science and Technology* 61 (7), 1410-1423.

Milojević, S. (2010b). Power law distributions in information science: making the case for logarithmic binning. *Journal of the American Society for Information Science and Technology* 61 (12), 2417-2425.

Mitzenmacher, M. (2004). A brief history of generative models for power law and lognormal distributions. *Internet Mathematics*, 1 (2), 226-251.

Newman, M. E. J. (2005). Power laws, Pareto distributions and Zipf's law. *Contemporary Physics*, 46 (5), 323-351.
53